\documentclass[submission,copyright,creativecommons]{eptcs}
\providecommand{\event}{HCVS 2023} 

\usepackage{amsmath}
\usepackage{iftex}
\usepackage{syntax}
\usepackage{semantic}
\usepackage{amsfonts}
\usepackage{listings}
\usepackage{fancyvrb}
\usepackage{float}
\usepackage{xcolor}

\ifpdf
  \usepackage{underscore}         
  \usepackage[T1]{fontenc}        
\else
  \usepackage{breakurl}           
\fi

\newcommand{\isint}[1] {
  \texttt{(\_ is anInt) #1)}}

\title{An Encoding for CLP Problems in SMT-LIB}

\author{Daneshvar Amrollahi
\institute{University of Tehran}
\email{d.amrollahi@ut.ac.ir}
\and
Hossein Hojjat
\institute{TeIAS, Khatam University\\ University of Tehran
}
\email{\quad hojjat@ut.ac.ir}
\and
Philipp R\"ummer
\institute{University of Regensburg\\
Uppsala University}
\email{philipp.ruemmer@ur.de}
}
\def\titlerunning{An Encoding of CLP Problems in SMT-LIB}
\def\authorrunning{D. Amrollahi, H. Hojjat \& P. R\"ummer}

\begin{document}
\maketitle

\begin{abstract}
The input language for today's CHC solvers are commonly the standard SMT-LIB format, borrowed from SMT solvers, and the Prolog format that stems from Constraint-Logic Programming (CLP). 
This paper presents a new front-end of the Eldarica CHC solver that allows inputs in the Prolog language.
We give a formal translation of a subset of Prolog into the SMT-LIB commands.
Our initial experiments show the effectiveness of the approach and the potential benefits to both the CHC solving and CLP communities.
\end{abstract}

\section{Introduction}\label{sec:introduction}

Over the last years, a growing number of solvers for Constrained Horn Clauses (CHC) have been developed; for instance, Spacer~\cite{DBLP:conf/cav/KomuravelliGC14}, Eldarica~\cite{8603013}, Golem~\cite{golem}, and RInGEN~\cite{ringen}. There are two main languages used to interface such solvers: the SMT-LIB language~\cite{BarFT-RR-17}, in which Horn clauses can either be expressed using quantified assertions, or using the rule-based notation that was introduced by Z3; and dialects of Prolog~\cite{ISO:1995:IIIe} as a language in Constraint-Logic Programming (CLP). 
The former language is designed primarily for machine-generated input to solvers, as it is simple to parse, strongly typed, and has unambiguous semantics.
The latter language is more concise and convenient for handwritten programs. 
There is, unfortunately, no exact match between the theories considered in CLP and SMT-LIB, and some Prolog language features have semantics that are not straightforward to model; for instance, the default absence of the ``occurs-check'' in equations.

This paper presents an ongoing effort to add a comprehensive Prolog/CLP front-end to the CHC solver Eldarica~\cite{8603013}. Eldarica has been able to process clauses in Prolog format since the beginning of its development; however, the existing Prolog front-end of Eldarica is restricted to the parsing of clauses over integers, and it is planned to replace it with a front-end with more extensive support for the different Prolog features through this project. To document the applied interpretation of Prolog, we define a formal translation of a fragment of Prolog to the SMT-LIB language. 
In the scope of this paper, we focus on three key features of Prolog: functions, defining a Herbrand universe of values; lists; and integer numbers as introduced by CLP($\mathbb{Z}$). 
We model the semantics of those language features using a mapping to SMT-LIB data-types (i.e., algebraic data-types with free constructors) and SMT-LIB integers.

\paragraph{Example}
To illustrate the complementarity of CHC and CLP, we start with a simple routing example including integer arithmetic, lists, and functions in Prolog.
Figure~\ref{fig:cities} shows a CLP($\mathbb{Z}$) program to compute the distance between cities. The program consists of 9~facts and 2~rules. 
A fact of the form \texttt{distance($X$, $Y$, $Z$)} states that the direct distance between cities $X$ and $Y$ is $Z$. The rule in line~12 implies that distance is symmetric.
The meaning of \texttt{path($X$, $Y$, $Z$, $L$)} is that there exists a path of length $Z$ from $X$ to $Y$ where $L$ represents the path as a list of points in the format \texttt{waypoint($C$, $D$)}. This shows that city $C$ lies on the path from $X$ to $Y$ with a distance of $D$ from the starting point, which is $X$. 
The fact \texttt{path(A, A, 0, [waypoint(A, 0)]).} states that for any city \texttt{A}, there is a path of length 0 to itself and the list presenting the path is the city \texttt{A} itself. 
Finally, the rule on line~15 implies that if there is path \texttt{N} of length \texttt{P} from \texttt{A} to \texttt{B}, 
and the direct distance from \texttt{B} to \texttt{C} is \texttt{Q} and \texttt{D} is equal to \texttt{P + Q}, then we have found a path from \texttt{A} to \texttt{C}, which is the previously found path (\texttt{N}) extended with the city~\texttt{C}. 
The query \texttt{path(tehran, munich, D, X), D \#< 40.} searches for a path from \texttt{munich} to \texttt{tehran} of length less than 40. 
A possible answer to this query is \texttt{D = 34} and \texttt{X = [waypoint(munich,34), waypoint(vienna,31), waypoint(tehran,0)]}. This means that there is a path from \texttt{tehran} to \texttt{munich} of distance \texttt{34}, with \texttt{vienna} being on the way with distance \texttt{31} from \texttt{tehran}. 

Note that this Prolog program, while intuitive, is also rather inefficient, and interpreting it using a CLP engine might lead to non-termination due the recursion on lines~12,15. 
The program can be rewritten to a more operationally oriented (and harder to read) set of clauses to be terminating and efficient, in particular preventing cyclic paths from being explored, and inlining the length bound to allow branches without solutions to be pruned. 
Alternatively, tabling~\cite{ChenWarren:JACM96} could be used for the predicate \texttt{path}\footnote{Which, however, led to an error in experiments with SWI-Prolog~\cite{wielemaker2010swiprolog}.}.

CHC solvers are generally less efficient than Prolog engines for solving constraint satisfaction problems.  However,
the abstraction techniques in CHC solvers are naturally able to cope with clauses written in declarative and otherwise inefficient style. They can easily find a path from \texttt{tehran} to \texttt{munich} of length \texttt{40} in Figure~\ref{fig:cities}. 
CHC solvers are also able to determine that no such path exists for lengths less than \texttt{34}, a task more challenging for at least some CLP solvers. More generally, CHC solvers are agnostic of the order of clauses, and the order of literals in clauses, and process clauses focusing more on their logical content than syntactic features.
In this sense, CHC solvers can be a useful debugging tool, and have a complementary performance profile to CLP systems. The goal of our work is to simplify the integration of CLP and CHC methods, by providing a translation of a subset of Prolog into SMT-LIB, the common input language of CHC solvers.

\begin{figure}
\centering
\begin{lstlisting}[frame=single, basicstyle=\ttfamily\footnotesize,numbers=left]
:- use_module(library(clpz)).  % or clpfd

distance(tehran,   vienna, 31).
distance(vienna,   paris,  10).
distance(vienna,   munich, 3).
distance(paris,    munich, 10).
distance(paris,    rome,   11).
distance(lausanne, rome,   6).
distance(lausanne, munich, 4).
distance(tehran,   paris,  42).

distance(A, B, D) :- distance(B, A, D).

path(A, A, 0, [waypoint(A, 0)]). 
path(A, C, D, [waypoint(C, D) | N]) :- path(A, B, P, N), distance(B, C, Q), 
                                        D #= P + Q. 
                                        
?- path(tehran, munich, D, X), D #< 40.
\end{lstlisting}  

\caption{Prolog Program for distance between cities in CLP($\mathbb{Z}$)}
    \label{fig:cities}
\end{figure}


\emph{Organization of the paper:} 
In Section~\ref{sec:prelim} we overview the CLP and the SMT-LIB input languages. 
Section~\ref{sec-translation} discusses translation rules for converting Prolog to SMT-LIB.
Section~\ref{sec:motivating-smt-lib} translates the introductory example to SMT-LIB.
We conclude the paper and present directions of future research in Section~\ref{sec:conc}.

\section{Preliminaries\label{sec:prelim}}

\subsection{Constraint Logic Programming (CLP)} 
Constraint Logic Programming (CLP)~\cite{Wallace2002}, first introduced by Jaffar and Lassez in 1987~\cite{10.1145/41625.41635}, is a programming paradigm that combines the benefits of constraint programming and logic programming. It enables the modeling of complicated real-world problems with variables and constraints, and uses logical and constraint reasoning to find a solution. CLP expresses the connections between variables as constraints and looks for a derivation of queries from given clauses.

Prolog is the main language of CLP, where the constraints are equations over the algebra of terms. CLP problems may in addition contain symbols with pre-defined meanings, defined by a \emph{theory}. CLP($\mathbb{Z}$) \cite{DBLP:conf/flops/Triska12} refers to CLP over the theory of integer arithmetic. Figure~\ref{fig:prolog-syntax} is a simplified grammar that is able to parse Prolog programs with functions, lists, and integer arithmetic. An extended version of it is used in our actual implementation.

\begin{figure}[]
{
\footnotesize

\begin{minipage}{.5\textwidth}
\begin{grammar}
    
<Database> ::= <Clause>\(^{*}\)

<Clause> ::= <Predicate> `.'
\alt <Predicate> `:-' <BodyItem>\(^{*}\) `.'
\alt `?-' <BodyItem>\(^{*}\) `.'

<BodyItem> ::= <Predicate>
\alt <Constraint>

<Predicate> ::= <Atom>
\alt <Atom> `(' <Term>\(^{*}\) `)'
\end{grammar}
\end{minipage}
\begin{minipage}{.5\textwidth}
\begin{grammar}

<Term> ::= <Variable>
\alt <Atom>
\alt <Atom> `(' <Term>\(^{*}\) `)'
\alt <List>
\alt <Integer>

<List> ::= `[' <Term>\(^{*}\) `]'
\alt `[' <Term>\(^{*}\) `|' <Term> `]'

\end{grammar}
\end{minipage}
}
\caption{A simplified grammar for describing the syntax of Prolog programs.}
\label{fig:prolog-syntax}
\end{figure}

As shown in Figure~\ref{fig:prolog-syntax}, every Prolog program consists of a set of \emph{Clause}s. A clause has a \emph{head} and a \emph{body}. The head can have zero or one predicates, and the body is a list of predicates and constraints. A clause is either a \emph{Fact}, a \emph{Rule}, or a \emph{Query}. For example, \texttt{man(tom)} is a fact, and \texttt{\mbox{friends(X, Y)} :- likes(X, Y), likes(Y, X)} is a rule with \texttt{friends(X, Y)} as the head and \texttt{likes(X, Y), likes(Y, X)} as the body. Queries always start with a \texttt{?-}. For instance, \texttt{?- likes(X, tom)} would search for an assignment for X such that \texttt{likes(X, tom)} can be derived from the set of given clauses. \emph{Term}s can be variables, atoms, compound terms, lists, and integers.

Structured data is represented using compound terms. A compound term consists of a function and a sequence of one or more sub-terms. A function is characterized by its name, which is an atom, and its arity. For instance, in the fact \texttt{man(father(claire))}, the term \texttt{father(claire)} is a compound term.

In Prolog, there are different kinds of built-in equality operators (\texttt{\#=, =, =:=, is}) with different semantics. 
The \texttt{=/2} predicate, which accepts two arguments, is used to ensure that its two arguments are syntactically equal through unification. For instance, the response to the query \texttt{?- father(X) = father(john)} will be positive as the sub-terms can be unified by setting \texttt{X = john}.
By default, Prolog does not apply any ``occurs-check'', however, which can sometimes cause non-termination. An example of this is \texttt{?- X = father(X)}. The unification algorithm decides to unify \texttt{X} with the right-hand-side, which is \texttt{father(X)}. But still there is an \texttt{X} left in this term; interpretation can get stuck in this loop of replacing \texttt{X} by \texttt{father(X)} and never terminate. 

\emph{List}s are terms representing sequences of elements; for instance, \texttt{[a, [b, c], 7]} is a list of terms. A non-empty list can be also be thought of as having two parts: \emph{head} as the first element of the list, and \emph{tail} as the remainder of the list. Both representations can be parsed using the grammar rules in Figure~\ref{fig:prolog-syntax}.

The body of the clauses in a Prolog program may also include \emph{constraint}s, in addition to predicates. An example of a constraint in CLP($\mathbb{Z}$), Constraint Logic Programming over the domain of integers, is \texttt{X + Y \#>= 3 * Z - 17}. Some other operators for constraints in CLP($\mathbb{Z}$) include \texttt{\#=, \#>, \#>=, \#<, \#=<, +, -, *, /, mod}. Throughout the rest of this paper, whenever we refer to constraints, we  refer to constraints in CLP($\mathbb{Z}$).

\subsection{SMT-LIB}

SMT-LIB, or the Satisfiability Modulo Theories Library~\cite{BarFT-RR-17}, is a standardized format for specifying logical formulas modulo various background theories. It is widely used in applications like software verification, hardware design, and automated reasoning, and is also used as a standard input format of CHC solvers. The SMT-LIB language supports a range of theories, including arithmetic, bit-vectors, and arrays, allowing users to express a wide variety of constraints.

An SMT-LIB script consists of a list of commands to be processed by an SMT solver. Important SMT-LIB commands are:
\begin{itemize}
    \item \texttt{(set-logic $L$)}: Set the logic, i.e., the combination of theories that will be used. For CHC solvers, typically $L = \texttt{HORN}$.
    \item \texttt{(declare-fun $f$ ($T_1~\ldots~T_n$) $T$)}: Declaration of a function or predicate~$f$ with the given argument types~$T_1~\ldots~T_n$ and result type~$T$. Types in SMT-LIB include the types provided by background theories (e.g., \texttt{Int} for integers, \texttt{Bool} for Booleans), as well as user-defined types. When interfacing CHC solvers, typically only predicates (i.e., Boolean-valued functions) are declared.
    \item \texttt{(declare-datatype $T$ ($C_1~\ldots~C_n$))}: Declaration of an algebraic data-type~$T$ with constructors~$C_1~\ldots~C_n$. Data-types can be recursive, and are the main modelling technique in SMT-LIB to represent structured data. For every constructor~$C$, the data-type declaration will also introduce a tester~\texttt{(\_ is $C$)} to identify terms constructed using $C$, as well as selectors for the arguments. An extended version of the command, \texttt{declare-datatypes}, exists to define several mutually recursive data-types. We will see various examples of algebraic data-types in this article.
    \item \texttt{(assert $\phi$)}: Assert a constraint~$\phi$, which can be formulated using the standard operators of first-order logic, the functions and predicates provided by background theories, and declared symbols. CHC solvers require that asserted constraints~$\phi$ are \emph{constrained Horn clauses,} which means that they fall into one of the following classes of formulas:
    \begin{center}
      \begin{tabular}{ll}
          Facts: & \texttt{($p$ $t_1~\ldots~t_k$)} \\
          Quantified facts: & 
          \texttt{(forall ($X_1~\ldots~X_n$) ($p$ $t_1~\ldots~t_k$))} \\
          Rules: & \texttt{(forall ($X_1~\ldots~X_n$) (=> (and $\psi_1~\ldots~\psi_m$) ($p$ $t_1~\ldots~t_k$)))} \\
          Queries: & \texttt{(forall ($X_1~\ldots~X_n$) (=> (and $\psi_1~\ldots~\psi_m$) false)}
      \end{tabular}
    \end{center}
    \item \texttt{(check-sat)}: Instruct the SMT solver to check whether the constraints asserted up to this point are satisfiable.
\end{itemize}

Throughout the paper, only a small subset of the SMT-LIB commands are used.
For a full definition of SMT-LIB, we refer to the standard documentation~\cite{BarFT-RR-17}.


\section{Translating CLP to SMT-LIB\label{sec-translation}}

In this section, we explain our translation from CLP problems in Prolog to SMT-LIB. We begin by a motivating example as a warm-up. After that, we present the translation rules for Prolog facts, rules, lists, and CLP($\mathbb{Z}$) constraints. 
We remark that our implementation does not explicitly translate input problems to SMT-LIB, but instead directly constructs clauses using Eldarica's internal data structures; the translation to SMT-LIB is presented for documentation purposes, and reflects the semantics that is applied.

\begin{figure}[t]

\begin{minipage}[t]{0.38\linewidth}
\begin{lstlisting}[]
(declare-datatype U (
    (claire)
    (father (father_1 U))
    )
)
\end{lstlisting}
\end{minipage}
\begin{minipage}[t]{0.27\linewidth}
\begin{lstlisting}[]
(declare-fun man 
    (U)
    Bool
)
\end{lstlisting}
\end{minipage}
\begin{minipage}[t]{0.25\linewidth}
\begin{lstlisting}[]
(assert
    (man (father claire))
)
\end{lstlisting}
\end{minipage}

\caption{SMT-LIB equivalent of \texttt{man(father(claire))}}
    \label{fig:claire}
\end{figure}

\subsection{A Motivating Example} \label{sec-motivating}

Consider the Prolog fact \texttt{man(father(claire))}. 
This fact consists of the function \texttt{father} applied to an atomic term \texttt{claire}. 
To model this term, one could consider declaring corresponding uninterpreted functions in SMT-LIB. However, with uninterpreted functions, \texttt{claire} and \texttt{father(claire)} might be assigned the same value, despite being syntactically different terms: the equation \texttt{(= claire (father claire))} is satisfiable in SMT-LIB. 
This is inconsistent with Prolog semantics, in which \texttt{claire} and \texttt{father(claire)} are different and non-unifiable terms, and it is guaranteed that they stand for distinct elements of the Herbrand universe. In addition, most CHC solvers do not support uninterpreted functions.

We therefore propose to treat \texttt{claire} and \texttt{father} as constructors of an algebraic data-type.  
To translate the mentioned fact to SMT-LIB, the first step is to introduce a new algebraic data-type for all of the terms occurring in a program. The data-type will be called \texttt{U}, standing for \emph{U}niversal. 
The constructors of \texttt{U} will be all atoms and functions appearing in the clauses. 
Figure~\ref{fig:claire} shows the full translation of the example. 
The constructor~\texttt{father} is unary, and its definition also adds a selector~\texttt{father_1} to retrieve the sub-term. 
We also define the predicate~\texttt{man}, as a Boolean function with a single argument of type \texttt{U}. 
Finally, we add the fact \texttt{man(father(claire))} to the set of our clauses using the \texttt{assert} command.
 Section~\ref{sec-adt} provides further details.

It has to be noted that our encoding of terms does not exactly model Prolog semantics.  As explained in Section~\ref{sec:prelim}, the unification algorithm of Prolog will not detect the non-unifiability of a query \texttt{?- X = father(X)} due to the missing occurs-check. In contrast, elements of SMT-LIB data-types are finite constructor terms, which implies that the equation~\texttt{(= X (father X))} is not satisfiable in SMT-LIB. However, we believe that the finite-term semantics is usually the intended semantics of a Prolog program. It remains to be investigated in future work whether there is a way to represent Prolog semantics more closely in SMT-LIB.

\begin{figure}[tb]
    \centering

    \[
    \mathit{collectFunctions}(t)=\begin{cases}
            \{(atom, 0)\}, & \text{if } t = \mathit{atom}\\
            \{(c,n)\} \cup \bigcup_{i = 1}^{i = n} \mathit{collectFunctions}(a_i), & \text{if } t = c(a_1, \cdots, a_n) \\
            \emptyset, & \text{otherwise}\\
		 \end{cases}
      \]
    
    \caption{Function $\mathit{collectFunctions}$ collecting functions in the terms}
    \label{fig:functions-collect}
\end{figure}

\begin{figure}[tb]
\centering
\begin{minipage}{0.36\linewidth}
\begin{lstlisting}[mathescape=true]   
(declare-datatype U
(
    ($f_1$ ($f_1^1$ U) ... ($f_1^{a_1}$ U))
    ...
    ($f_n$ ($f_n^1$ U) ... ($f_n^{a_n}$ U))
)
\end{lstlisting}
\end{minipage}

\caption{Definition of the universal type \texttt{U} for functions~$\{ (f_1, a_1), \ldots, (f_n, a_n) \}$}
\label{fig:univ-def}
\end{figure}

\subsection{Algebraic Data Types (ADTs)} \label{sec-adt}

We define a single algebraic data-type~\texttt{U} for representing the type of all terms in a program. 
The constructors of \texttt{U} will be all functions and atoms appearing in the clauses. 
As this global type \texttt{U} is defined only once, we need to iterate through all clauses and collect the functions and atoms appearing in them. 
We define a recursive function responsible for this matter in Figure~\ref{fig:functions-collect}.
This function is applied to a term~$t$, and recursively collects the atoms and functions occurring in $t$, as well as their arity. For instance, the value of $\mathit{collectFunctions}(a(X, b, c(d, Y, e)))$ is $\{
    (e, 0), (d, 0), (c, 3), (b, 0), (a, 3)
\}$.

Given that the set collected using \(\mathit{collectFunctions}\) is \( \{ (f_1, a_1), \ldots, (f_n, a_n) \} \), the universal algebraic data type \texttt{U} should be defined as in Figure~\ref{fig:univ-def}.

\begin{figure}[tb]
        \begin{equation*}
        \begin{array}{c@{\qquad}c}
        \inference
        {}
        {a \; \triangleright \; a \; : \emptyset}
        [a is an atom]
             & 
        \inference
        {}
        {V \; \triangleright \; V \; : \emptyset}
        [V is a variable]
        \\[5ex]
        \multicolumn{2}{c}{
        \inference
        {\{t_i \; \triangleright \; t_i^\prime \; : \; \Phi_i\}_{i = 1}^{n}}
        {\texttt{f($t_1$,...,$t_n$)} \; \triangleright \; \texttt{(f $t_1^\prime$ ... $t_n^\prime$)} \; : \; \bigcup_{i = 1}^{n} \Phi_i}
        [\texttt{f} is a function or predicate]
        }
        \\[5ex]
        \inference 
        {s \; \triangleright \; s^\prime \; : \; \Phi_1 \quad t \; \triangleright t^\prime \; : \; \Phi_2}
        { \texttt{$s$ = $t$} \; \triangleright \; \texttt{(= $s$\(^\prime\) $t$\(^\prime\))} \; : \; \Phi_1 \cup \Phi_2  }[]
        &    
        \inference 
        {s \; \triangleright \; s^\prime \; : \; \Phi}
        { \texttt{ \textbackslash+$s$} \; \triangleright \; \texttt{(not $s$\(^\prime\))} \; : \; \Phi}[]
        \\[5ex]
        \multicolumn{2}{c}{
        \inference 
        {s \; \triangleright \; s^\prime \; : \; \Phi_1 \quad t \; \triangleright \; t^\prime \; : \; \Phi_2}
        { \texttt{$s$ =\textbackslash= $t$} \; \triangleright \; \texttt{(not (= $s$\(^\prime\) $t$\(^\prime\)))} \; : \; \Phi_1 \cup \Phi_2}[]}
                \end{array}
        \end{equation*}

    \caption{Basic rules for translating Prolog to SMT-LIB}
    \label{fig:base-cases}
\end{figure}

\begin{figure}
\centering
\begin{minipage}[t]{0.37\linewidth}
\begin{lstlisting}[mathescape=true]
(declare-fun $p$
    (U ... U)
    Bool
)
\end{lstlisting}
\end{minipage}
~
\begin{minipage}[t]{0.6\linewidth}
\begin{lstlisting}[mathescape=true]
(assert
    (forall ( ($X_1$ U) ... ($X_m$ U) )
        (=> (and $\Phi_1$ ... $\Phi_n$) ($p$ $t_1^\prime$ ... $t_n^\prime$))
    ) 
)
\end{lstlisting}
\end{minipage}

\caption{Translation of a fact~\(p(t_1, \ldots, t_n)\). We assume that the elements of a set~$\Phi_i$ are implicitly conjoined.}
\label{fig:facts}
\end{figure}

\subsection{Translation of Facts} \label{sec-facts}

Section~\ref{sec-motivating} shows an example of translating a fact to SMT-LIB. We now introduce the general case of this translation, and add rules for lists and integer arithmetic
in Sections~\ref{sec-lists}~and~\ref{sec-ints}.
We refer to  triplets of the form $\texttt{s} \; \triangleright \; \texttt{s}^\prime \; : \; \Phi$  as \emph{translation judgement}s. The meaning of a judgement is that the Prolog term~\texttt{s} can be translated to an SMT-LIB term~\texttt{s}$^\prime$ under a set of side conditions~$\Phi$. Side conditions are mainly needed to capture typing requirements. For instance, Prolog (CLP($\mathbb{Z}$)) raises an error when encountering the expression \texttt{a + b}, where \texttt{a} or \texttt{b} are not integers. Thus, when translating \texttt{a + b} to SMT-LIB, we will later add side conditions that ensure the correct type of \texttt{a} and \texttt{b}. More details concerning the use of side conditions are given in Sections~\ref{sec-lists}~and~\ref{sec-ints}.

Figure~\ref{fig:base-cases} shows the basic translation rules. Each rule derives a translation judgement, the conclusion under the bar, from premises shown above the bar. The rules in Figure~\ref{fig:base-cases} are mostly self-explanatory, and recursively translate a given term or constraint to SMT-LIB. The rules assume, for sake of presentation, that atoms, variables, and functions are translated to SMT-LIB symbols with the same name.

The SMT-LIB translation of a Prolog fact~\(p(t_1, \ldots, t_n)\) is shown in Figure~\ref{fig:facts} in terms of expressions \(t_1^\prime, \ldots, t_n^\prime \) and side conditions \(\Phi_1, \ldots, \Phi_n \), which are defined as follows:  
\begin{itemize}
    \item For all $i \in \{1, \ldots, n\}$ it holds that \( t_i \; \triangleright \; t_i^\prime \; : \; \Phi_i \).
    \item \(X_1, \ldots, X_m\) are all the variables appearing in the terms \(t_1, \ldots, t_n\).
\end{itemize}
The fact is captured using an SMT-LIB assertion in Figure~\ref{fig:facts}, listing the side conditions required by the translation as assumptions.

%

\begin{figure}
\centering
\begin{minipage}[t]{0.37\linewidth}
\begin{lstlisting}
(declare-fun likes 
    (U U)
    Bool
)

(declare-fun friends 
    (U U)
    Bool
)
\end{lstlisting}
\end{minipage}
~
\begin{minipage}[t]{0.5\linewidth}
\begin{lstlisting}
(assert 
    (forall ( (X U) (Y U) )
        (=> 
            (and (likes X Y) (likes Y X)) 
            (friends X Y) 
        ) 
    ) 
)
\end{lstlisting}
\end{minipage}
\caption{Translation of a rule in SMT-LIB}
\label{fig:friends}
\end{figure}

\begin{figure}
\centering
\begin{minipage}{0.55\linewidth}
\begin{lstlisting}[mathescape=true]
(assert
    (forall ( ($X_1$ U) ... ($X_k$ U) )
        (=> (and $\psi_1^\prime \; \ldots \; \psi_m^\prime$ $\Phi_0$ ... $\Phi_{m}$) $\psi'_0$)
    ) 
)
\end{lstlisting}
\end{minipage}

\caption{Translation of a rule~\( \psi_0 \; \texttt{:-} \; \psi_1, \ldots, \psi_m \)}
\label{fig:rules}
\end{figure}

\subsection{Translation of Rules}

‌‌Before presenting the translation of rules, 
we begin with a concrete example, using the Prolog rule
\texttt{friends(X, Y) :- likes(X, Y), likes(Y, X)}. This rule will be
turned into a universally quantified implication in SMT-LIB. Given that we have already defined our universal data-type~\texttt{U} as explained in Section~\ref{sec-adt}, all we need is to declare functions in SMT-LIB for \texttt{likes} and \texttt{friends}, and create the rule using the universal quantifier. The full translation is illustrated in Figure~\ref{fig:friends}.

The general schema for expressing a rule~\( \psi_0 \; \texttt{:-} \; \psi_1, \ldots, \psi_m \) in SMT-LIB is shown in Figure~\ref{fig:rules}, assuming that:
\begin{itemize}
    \item For all $i \in \{0, \ldots, m\}$ it holds that \( \psi_i \; \triangleright \; \psi_i^\prime  \; : \; \Phi_i\).
    \item \(X_1, \ldots, X_k\) are all the variables appearing in the rule.
\end{itemize}
The assertion representing the rule in Figure~\ref{fig:rules} has a similar shape as the assertion for facts in Figure~\ref{fig:facts}.



\begin{figure}
\centering
\begin{minipage}{0.5\linewidth}
\begin{lstlisting}
(declare-datatypes () (
        (U 
            (aList (theList L))
            ...
        )
        (L 
            nil 
            (cons (head U) (tail L))
        )
    ) 
)
\end{lstlisting}
\end{minipage}

\caption{Algebraic data-types including lists}
    \label{fig:LU}
\end{figure}

\begin{figure}[t]
\centering
\begin{minipage}{0.75\linewidth}
\begin{lstlisting}[mathescape=true, basicstyle=\ttfamily\footnotesize]    
list_concat([],L,L).
list_concat([X1|L1],L2,[X1|L3]) :- list_concat(L1,L2,L3).
\end{lstlisting}
\end{minipage}

\caption{A Prolog program concatenating lists}
\label{fig:ex-concat}
\end{figure}

\begin{figure}
\begin{lstlisting}[]
(declare-fun list_concat (U U U) Bool)

(assert (forall ((X U)) (list_concat (aList nil) X X)))

(assert (forall ((X1 U) (L1 U) (L2 U) (L3 U))
    (=> (and (list_concat L1 L2 L3) ((_ is aList) L1) ((_ is aList) L3))
        (list_concat
          (aList (cons X1 (theList L1))) L2 (aList (cons X1 (theList L3))))))
)
\end{lstlisting}

\caption{SMT-LIB representation of the program in Figure~\ref{fig:ex-concat}}
    \label{fig:ex-list}
\end{figure}

\begin{figure}[tb]
   \begin{gather*}
    \inference 
        {}
        {
        \texttt{[]} \; \triangleright \; \texttt{(aList nil)} \; : \emptyset
        }
        []
    \\[4ex]
    \inference 
        {\{t_i \; \triangleright \; t_i^\prime \; : \; \Phi_i\}_{i = 1}^{n}}
        { \texttt{ [$t_1$,...,$t_n$] } \triangleright  
        \texttt{
        (aList (cons $t_1^\prime$ (cons $t_2^\prime$ (... cons $t_n^\prime$ nil))...))}
        \triangleright \; \bigcup_{i = 1}^{n} \Phi_i
        }
        []
    \\[4ex]
    \inference 
        {h \; \triangleright \; h^\prime \; \triangleright \; \Phi_1 \quad t \; \triangleright \; t^\prime \; \triangleright \; \Phi_2}
        {
        \texttt{[$h$|$t$]} \; \triangleright \; 
        \texttt{
        (aList (cons $h^\prime$ (theList $t'$)))
        } \; : \; \Phi_1 \cup \Phi_2 \cup \{\texttt{((\_ is aList) $t^\prime$)}\}
        }
        []
   \end{gather*}

    \caption{Translation rules for lists}
    \label{fig:list-rules}
\end{figure}

\subsection{Translation of Lists} \label{sec-lists}

We now add lists to the picture. The absence of static typing in Prolog makes it a-priori impossible to tell whether a Prolog variable will represent a list or a term constructed using some function (or possibly both). We therefore include lists as one case in our universal data-type~\texttt{U}, and ensure the correct typing of terms through side-conditions in our translation rules.

Lists can be viewed as an algebraic data-type (call it \texttt{L}) with two constructors \texttt{nil} and \texttt{cons}. The constructor \texttt{nil} has zero arguments and represents empty lists, and \texttt{cons} has two arguments: one referring to the head of the list (a term of type \texttt{U}), and one of type \texttt{L} referring to the tail of the list.

To implement lists in SMT-LIB, it is therefore natural to declare a new data-type~\texttt{L} with the \texttt{nil} and \texttt{cons} constructors. This new data-type will have a mutual dependency on our universal data-type \texttt{U}, as defined before, and therefore has to be declared along with \texttt{U} as shown Figure~\ref{fig:LU}. Note that in Figure~\ref{fig:LU}, there may be other constructors for \texttt{U} coming from the collected functions in Section~\ref{sec-adt}, denoted by three dots. The \texttt{aList} constructor is introduced to wrap terms of type~\texttt{L} as a term of the universal type~\texttt{U}. 
An example of a Prolog program involving lists is presented in Figure~\ref{fig:ex-concat}. The translation of the program in Figure~\ref{fig:ex-concat} to SMT-LIB is presented in Figure ~\ref{fig:ex-list}.

A set of inference rules for translating lists in Prolog to SMT-LIB is presented in Figure~\ref{fig:list-rules}. The first rule in Figure~\ref{fig:list-rules} translates the empty list. The second rule represents the translation of a list with explicitly enumerated elements. In this case, the list can be constructed in SMT-LIB as a simple iterated application of \texttt{cons}. The final rule directly matches the functional definition of a list. A list in Prolog with $h$ as the head, and $t$ as the tail, can be constructed using the \texttt{cons} constructor in SMT-LIB. This construction is type-correct only if $t$ is again a list; thus, the rule adds the side condition \texttt{((\_ is aList) $t^\prime$)}, and unwraps the tail using the \texttt{theList} selector. In all the rules in Figure~\ref{fig:list-rules}, the result is finally wrapped using the \texttt{aList} constructor and turned into a \texttt{U}-term.

\begin{figure}
\centering
\begin{minipage}{0.45\linewidth}
\begin{lstlisting}
(declare-datatypes () (
        (U 
            (anInt (theInt Int))
            ...
        )
    ) 
)
\end{lstlisting}
\end{minipage}

\caption{Modified declaration of the universal ADT for integer arithmetic}
    \label{fig:U-ints}
\end{figure}

\begin{figure}
   \begin{gather*}
    \inference 
        {s \; \triangleright \; s^\prime \; : \; \Phi_1 \quad t \; \triangleright \; t^\prime \; : \; \Phi_2}
        { s \star t \; \triangleright \; \texttt{(anInt ($\mathit{op}$ (theInt $s^\prime$) (theInt $t^\prime$)))} \; : \; \Phi_1 \cup \Phi_2 \cup \{ \isint{$s^\prime$}, \isint{$t^\prime$} \} }
        []
    \\[1ex]
    \text{where~~}
    (\star, \mathit{op}) \in \{\texttt{(+,+), (-,-),(*,*),(/,div),(mod,mod)}\}
    \\[4ex]
    \inference 
        {s \; \triangleright \; s^\prime \; : \; \Phi_1 \quad t \; \triangleright \; t^\prime \; : \; \Phi_2}
        { s \circ t \; \triangleright \; \texttt{($\mathit{op}$ (theInt $s^\prime$) (theInt $t^\prime$))} \; : \; \Phi_1 \cup \Phi_2 \cup \{ \isint{$s^\prime$}, \isint{$t^\prime$} \} }
        []
    \\[1ex]
    \text{where~~}
    (\circ, \mathit{op}) \in \{\texttt{(\#=,=), (\#>,>),(\#>=,>=),(\#<,<),(\#=<,<=)}\}
    \\[4ex]
    \inference
    {}
    {I \; \triangleright \; \texttt{(anInt $I$)} \; : \; \emptyset}
    [I is a decimal non-negative integer]
    \\[4ex]
    \inference
    {}
    {\texttt{-}I \; \triangleright \; \texttt{(anInt (- $I$))} \; : \; \emptyset}
    [I is a decimal non-negative integer]
    \end{gather*}

    \caption{Translation rules for expressions and contraints in CLP($\mathbb{Z}$)}
    \label{fig:clpz-rules}
\end{figure}

\subsection{Integer Arithmetic} \label{sec-ints}

We finally introduce inference rules for converting constraints in the background theory of integer arithmetic (e.g., \texttt{2*X + 7 \#> Y}) to their corresponding expressions in SMT-LIB. To represent integers, similarly as with lists we add a new constructor to \texttt{U}, so that we can wrap integers into a term of type \texttt{U}. We call this new constructor \texttt{anInt}, and define it in Figure~\ref{fig:U-ints}. We once again note that \texttt{U} is a universal type and may include other constructors.

The translation rules for integer expressions and Boolean constraints over them are presented in Figure~\ref{fig:clpz-rules}. The first rule translates integer-valued functions by recursively translating the sub-terms, unwrapping the result, and applying the corresponding SMT-LIB function. The result is finally wrapped again using \texttt{anInt}. The second rule performs the corresponding translation for integer predicates. The other two rules take care of the translation of integer literals.

The translation can be modified easily to model bounded integers, which are used by some Prolog implementations. In this case, instead of \texttt{Int} in Figure~\ref{fig:U-ints}, a bit-vector type like \texttt{(\_ BitVec 64)} has to be used, and in Figure~\ref{fig:clpz-rules} the corresponding bit-vector operations have to be applied. Bit-vector support in CHC solvers is much less mature than support for mathematical integers, however.

As an example, consider the expression \texttt{X \#> 7}. According to the translation rule for variables in Figure~\ref{fig:base-cases}, $\texttt{X} \; \triangleright \; X \; : \; \emptyset $, and according to the translation rule for integers in Figure~\ref{fig:clpz-rules}, $\texttt{7} \; \triangleright \; 
 \texttt{(anInt 7)} \; : \; \emptyset$. Finally, according to the first rule in Figure~\ref{fig:clpz-rules} for translating expressions in integer arithmetic, the whole expression gets translated to \texttt{(anInt (+ (theInt X) (theInt (anInt 7))))}, with the side constraint that \texttt{X} is an integer.

\section{An SMT-LIB Encoding of the Motivating Example}
\label{sec:motivating-smt-lib}

Combining the different Prolog features that were discussed,
Figure~\ref{fig:cities-smt} gives the SMT-LIB encoding of the Prolog program for computing paths between cities from Section~\ref{sec:introduction}. All the commands used in this encoding can be obtained using the translation rules explained throughout this article.  
The encoding begins by declaring the universal data-type \texttt{U}. Its constructors include all the atoms and functions that appear in the clauses, in addition to \texttt{anInt} and \texttt{aList}. After that, the data-type \texttt{L} is declared for lists, as explained in Section~\ref{sec-lists}. The SMT-LIB script continues by declaring the functions \texttt{distance} and \texttt{path} appearing in the Prolog program clauses. The rest of the encoding are assertions corresponding to the Prolog rules and facts, and finally the assertion as a clause with head \texttt{false}.

A CHC solver like Eldarica~\cite{8603013} can derive the status \texttt{unsat} for this SMT-LIB script, which implies that the clauses are contradictory. This can be interpreted as the negation of the last clause being derivable from the other clauses. A CHC solver could also discover the path discussed in Section~\ref{sec:introduction}.

When tightening the length bound to \texttt{(< (theInt D) 34)}, the problem becomes satisfiable, which can again be verified using a CHC solver.

\begin{figure}
\centering
\begin{lstlisting}[frame=single, basicstyle=\ttfamily\footnotesize,numbers=left]
(declare-datatypes () (
        (U
            (anInt (theInt Int))
            (aList (theList L))
            (waypoint (waypoint_1 U) (waypoint_2 U))
            tehran vienna paris munich rome lausanne
        )
        (L
            nil 
            (cons (head U) (tail L))
        )  
    )
)
(declare-fun distance (U U U) Bool)
(declare-fun path (U U U U) Bool)

(assert (distance tehran   vienna (anInt 31)))
(assert (distance vienna   paris  (anInt 10)))
(assert (distance vienna   munich (anInt 3)))
(assert (distance paris    munich (anInt 10)))
(assert (distance paris    rome   (anInt 11)))
(assert (distance lausanne rome   (anInt 6)))
(assert (distance lausanne munich (anInt 4)))
(assert (distance tehran   paris  (anInt 42)))
(assert
    (forall ( (A U) (B U) (D U) )
        (=> (distance B A D) (distance A B D))
    )
)
(assert
    (forall ( (A U) )
        (path A A (anInt 0) (aList (cons (waypoint A (anInt 0)) nil)))
    )
)
(assert
    (forall ( (A U) (B U) (C U) (D U) (N U) (P U) (Q U) ) 
        (=>
            (and
                (path A B P N) (distance B C Q)
                (= D (anInt (+ (theInt P) (theInt Q))))
                ((_ is aList) N) ((_ is anInt) P) ((_ is anInt) Q)
            )
            (path A C D (aList (cons (waypoint C D) (theList N)))) 
        )
    )
)
(assert
    (forall ( (D U) (X U) ) 
        (=>
            (and (path tehran munich D X) (< (theInt D) 40) ((_ is anInt) D))
            false
        )
    )
)

(check-sat)
\end{lstlisting}

\caption{The SMT-LIB encoding of the motivating example introduced in the introduction}
    \label{fig:cities-smt}
\end{figure}

\section{Conclusion\label{sec:conc}}

We have presented work towards a new CHC solver front-end that allows input in Prolog format, bridging the gap between Prolog/CLP semantics and SMT-LIB. We are in the process of implementing the defined translation from Prolog to SMT-LIB in our Horn solver~Eldarica~\cite{8603013}, with the goal of achieving good coverage of the Prolog and CLP features.

The translation defined in this paper should be seen as a starting point, as there are several aspects that require further work, more research, or more discussion in the communities:
\begin{itemize}
    \item We have only shown the translation rules for some of the most important CLP($\mathbb{Z}$) operators. 
    We believe that many other theories and constraints can be handled in a similar fashion.
    \item We keep typing constraints dynamic, and this way stay close to the actual Prolog semantics. In terms of the efficiency of CHC solvers on the translated program, of course, it could be beneficial to perform some amount of type inference upfront. This way, one could translate integer variables in Prolog to native SMT-LIB \texttt{Int} variables, etc. However, CHC solvers with support for algebraic data-types tend to perform such type inference themselves, so that the payoff from being more clever during the translation is unclear.
    \item Our translation includes all typing constraints as assumptions (Figures~\ref{fig:facts} and \ref{fig:rules}), i.e., the well-typedness of a Prolog program is assumed but not verified. 
    It is not entirely obvious how correct typing should be asserted in the SMT-LIB representation, however.    
    In the list example in Figure~\ref{fig:ex-concat}, for instance, note that the first clause implies that the second and third argument of \texttt{list\_concat} can be terms of any kind, whereas the second clause relies on the third argument being a list. The clauses therefore entail ill-typed statements like \texttt{list\_concat([X|[]], 42, [X|42]])}. A CLP engine performing backward chaining will, however, not run into any typing errors.
    \item There are several aspects of Prolog semantics that are challenging in a translation to SMT-LIB. Those include, in particular, the \emph{missing occurs-check} in equations, as well as \emph{cuts.} It is unclear whether those features can or should be translated faithfully to SMT-LIB semantics.
\end{itemize}

Finally, it will be interesting to evaluate solver performance for the different design choices in the translation, for different CHC solvers, and to compare to the performance of state-of-the-art CLP engines. We have not done such a comparison yet due to the preliminary state of the implementation of the front-end. While we generally assume CLP to be significantly more efficient on classical constraint satisfaction problems than CHC, there might also be areas in which the abstraction-based algorithms used in CHC solvers have advantages.

\nocite{*}
\bibliographystyle{eptcs}
\bibliography{generic}
\end{document}